\documentclass{aa}
\usepackage{times}
\usepackage{graphicx}
\usepackage{caption}
\usepackage{txfonts}
\usepackage{natbib}
\usepackage{appendix}
\usepackage{xcolor}
\usepackage{footnote}

\begin{document} 
   \title{Jupiter's inhomogeneous envelope}

   \author{Y. Miguel\inst{1,2}
          \and
          M. Bazot\inst{3,4}
          \and
          T. Guillot\inst{5}
          \and
          S. Howard\inst{5}
          \and 
          E. Galanti\inst{6}
          \and 
          Y. Kaspi\inst{6}
          \and 
          W. B. Hubbard\inst{7}
          \and 
          B. Militzer\inst{8}
          \and 
          R. Helled\inst{9}
          \and
          S. K. Atreya\inst{10}
          \and 
          J. E. P. Connerney\inst{11,12}
          \and 
          D. Durante\inst{13}
          \and 
          L. Kulowski\inst{14}
          \and 
          J. I. Lunine\inst{15,16}
          \and 
          D. Stevenson\inst{17}
          \and 
          S. Bolton\inst{18}
          }

   \institute{SRON Netherlands Institute for Space Research , Niels Bohrweg 4, 2333 CA Leiden, the Netherlands,
              \email{ymiguel@strw.leidenuniv.nl}
         \and
             Leiden Observatory, University of Leiden, Niels Bohrweg 2, 2333CA Leiden, The Netherlands,
             \and
             Heidelberg Institute for Theoretical Studies (HITS gGmbH), Schloss-Wolfsbrunnenweg 35,69118 Heidelberg, Germany
             \and
             CITIES, NYUAD Institute, New York University Abu Dhabi, PO Box 129188, Abu Dhabi, United Arab Emirates,
             \and
             Université Côte d Azur, OCA, Lagrange CNRS, 06304 Nice, France,
             \and
             Department of Earth and Planetary Sciences,Weizmann Institute of Science, Rehovot 76100, Israel,
             \and
             Lunar and Planetary Laboratory, University of Arizona, Tucson, AZ 85721, USA,
             \and
             Department of Earth and Planetary Science, University of California, Berkeley, CA 94720, USA,
             \and
             Institute for Computational Science, University of Zurich, Winterthurerstr. 190, CH8057 Zurich, Switzerland,
             \and
             University of Michigan, Climate and Space Sciences and Engineering, Ann Arbor, MI 48109, USA,
             \and
             Space Research Corporation, Annapolis, MD 21403, USA,
             \and
             NASA Goddard Space Flight Center, Greenbelt, MD 20771, USA,
             \and
             Department of Mechanical and Aerospace Engineering, Sapienza University of Rome, Italy,
             \and
             Department of Earth and Planetary Sciences, Harvard University, Cambridge, Massachusetts, USA,
             \and
             Department of Astronomy, Cornell Unihttps://www.overleaf.com/projectversity, 122 Sciences Drive, Ithaca, NY 14853, USA,
             \and
             Carl Sagan Institute, Cornell University, 122 Sciences Drive, Ithaca, NY 14853, USA,
             \and
             Division of Geological and Planetary Sciences, California Institute of Technology, Pasadena, California 91125, USA,
             \and
             outhwest Research Institute, San Antonio, TX 78238, USA,
             }
             \date{}

   \abstract
   {While Jupiter's massive gas envelope consists mainly of hydrogen and helium, the key to understanding Jupiter's formation and evolution lies in the distribution of the remaining (heavy) elements. Before the Juno mission, the lack of high-precision gravity harmonics precluded the use of statistical analyses in a robust determination of the heavy-elements distribution in Jupiter's envelope.}
   {In this paper, we assemble the most comprehensive and diverse collection of Jupiter interior models to date and use it to study the distribution of heavy elements in the planet's envelope.}
   {We apply a Bayesian statistical approach to our interior model calculations, reproducing the Juno gravitational and atmospheric measurements and constraints from the deep zonal flows.}
   {Our results show that the gravity constraints lead to a deep entropy of Jupiter corresponding to a 1 bar temperature 5-15 K higher than traditionally assumed. We also find that uncertainties in the equation of state are crucial when determining the amount of heavy elements in Jupiter's interior.
   Our models put an upper limit to the inner compact core of Jupiter of 7 M$_{\rm Earth}$, independently on the structure model (with or without dilute core) and the equation of state considered. Furthermore, we robustly demonstrate that Jupiter's envelope is inhomogenous, with a heavy-element enrichment in the interior relative to the outer envelope. This implies that heavy element enrichment continued through the gas accretion phase, with important implications for the formation of giant planets in our solar system and beyond.}
   {}
   
   \maketitle
   
\section{Introduction}
The standard model of Jupiter formation starts with the accretion of solids followed by
a rapid gas accretion phase wherein Jupiter's hydrogen/helium envelope is captured from the primitive solar nebula. In this scenario, the dominant size of the accreted solids could be either kilometer-sized planetesimals \citep{po96} or centimeter-sized pebbles \citep{la14}, with dramatic implications for formation time-scales and the distribution of heavy elements in Jupiter's envelope \citep{ve20,va18}. In the planetesimal-driven scenario, the accretion of solids continues through the gas accretion phase and stops when all of the planetesimals in the planet's vicinity have been accreted. The fragmentation and ablation of these solid planetesimals cause a non-homogenous distribution of heavy elements in the envelope \citep{al18}. In contrast, in the pebble-driven scenario, fast orbital decay of pebbles caused by gas drag provides a continuous resupply of solid material that enriches the growing planet \citep{ormel21}. Nevertheless, the supply stops once the so-called pebble-isolation mass is reached \citep{la14}, after which only gas accretion continues unless or until pebbles grow to planetesimal size. Here we seek to distinguish between these scenarios, analysing the internal structure of Jupiter today and determining whether Jupiter's envelope harbours a non-homogeneous distribution of heavy elements. We explore the possibility of differentiation of the heavy elements in the envelope, testing whether this differentiation is a natural outcome of the set of models that reproduce all observational constraints, including those provided by the recent Juno observations.

\section{Methods}
\subsection{Jupiter's structure: 3-layer and dilute-core models}
\begin{figure}[h]
  \begin{center}
\includegraphics[angle=0,width=0.55\textwidth]{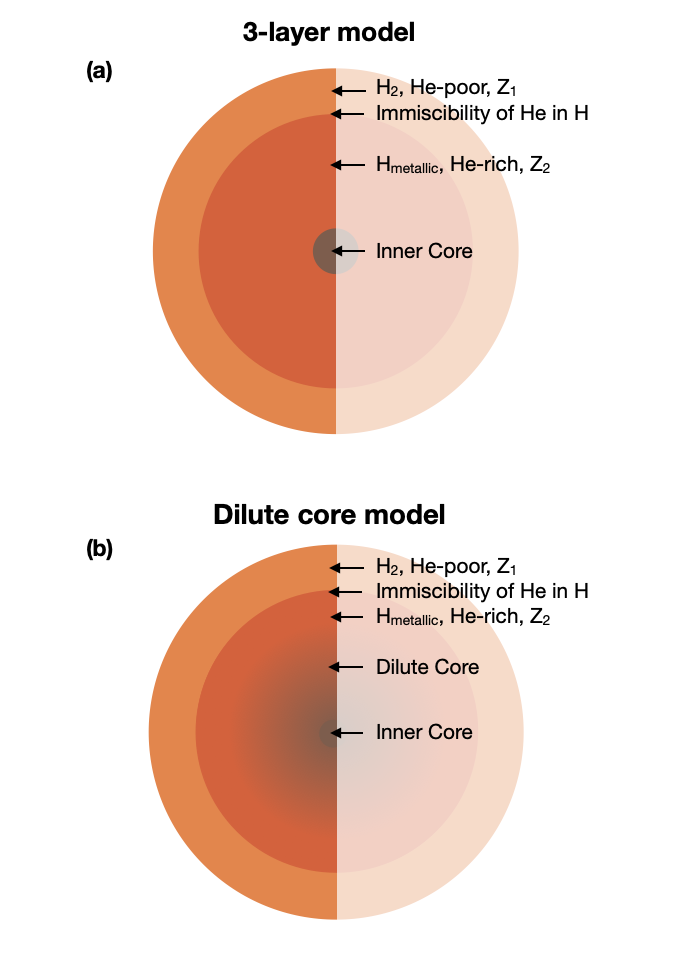}
 \end{center}
  \caption{Schematic view of the Jupiter structure models used in this paper. Panel (a): 3-layer model. Panel (b): Dilute core model. The different regions are indicated with arrow.} \label{Fig1}
\end{figure}

We construct two contrasting sets of Jupiter interior models and explore the largest possible ensemble of realistic models to date that represent the interior of Jupiter. The first set of models is constructed using a 3-layer Jupiter model where Jupiter's interior consists of an outer H$_2$-dominated region, an intermediate H$_{\rm metallic}$-dominated region (e.g. \cite{hu16}), and a core made of 100\% heavy elements (Fig. \ref{Fig1}a). The second set of models is built using 3 layers and adding a dilute core: a region above the inner core where the H and He of the envelope are gradually mixed with a greater abundance of heavy elements \citep{wa17} (Fig. \ref{Fig1}b). For the latter models we characterise the dilute core with three parameters: the maximum mass mixing ratio of heavy elements in the diluted core region ($Z_{\rm dilute}$), a parameter that controls the extent of the dilute core in terms of mass ($m_{\rm dilute}$) and a parameter controlling how steep is the heavy elements gradient ($\delta m_{dil}$). In the end, the heavy elements mass mixing ratio between the inner core outer boundary and the helium transition is given by :
\begin{equation}
\boldsymbol{Z} = Z_2 + \frac{Z_{\rm dilute}-Z_2}{2}\left[ 1 - {\rm erf}\left(\frac{m-m_{\rm dilute}}{\delta m_{dil}}\right)\right]
\end{equation}
with $\delta m_{dil}=0.075$. We tested possible variations of this parameter and found that the results of the calculation are not sensitive to the choice of $\delta  m_{dil}$. In addition, for all the models, we allow for an increase in the interior entropy in the helium-demixing region by including a temperature jump ($\Delta T_{\rm He}$) following \cite{hu16}. A posteriori examination of the models show that this region is convectively stable for values of this parameter lower than about 2000\,K \citep{gu18}. The outermost layer of the envelope in both sets of models is separated from the inner layer of the envelope by the transition to a region of immiscibility of helium in hydrogen ("helium rain"). The precise location of this region is not well known, but numerical calculations \citep{mo13,scr18} suggest that the immiscibility should be located  between $\sim 0.8$ and $\sim 3$ Mbar, and lab experiments indicate that it may extend even deeper into the planet \citep{br21}. In our models, we assume a Gaussian distribution taking these constraints into account. We run Markov chain Monte Carlo (MCMC) calculations for the two scenarios using a Bayesian statistical model \citep{ba12}, wherein calculations are concentrated around the relevant regions of the parameter space. This allows millions of solutions (models that fit Jupiter's observational constraints) to be evaluated with a feasible computational cost. See the Appendix for more details on the Bayesian approach. 

\subsection{Details on each Jupiter interior model}
We calculate each Jupiter interior structure model  with the code CEPAM \citep{cepam95,gu18}, that solves the equations of hydrostatic equilibrium, mass and energy conservation, and energy transport. We perform multiple sets of MCMC runs using different equations of state. Because Jupiter is primarily composed of hydrogen and helium, the equations of state of these two elements are fundamental when modelling its interior structure \citep{mi16}. Several equations of state have been published in the past few years and many uncertainties still remain \citep{mazevet20}. Therefore, our calculations are repeated using several proposed equations of state to avoid a bias in our results. We use MH13-H \citep{MH13,mi16}, CMS19-H \citep{cms19} and MLS21-H \citep{mazevet20} for the equation of state of hydrogen, and for helium we use SCVH95-He \citep{SCvH95} and CMS19-He \citep{cms19}. For the heavy elements, we use the equations of state of a mixture of silicates (dry sand) and the one of water in SESAME \citep{SESAME}. 

Jupiter's adiabat is prescribed by its entropy or, equivalently, by its temperature at the 1 bar level ($T_{\rm 1bar}$). Voyager radio-occultations \citep{Lindal1981} initially indicated $T_{\rm 1bar}=165 \pm5$\,K. In situ measurements from the Galileo probe measured, in a so-called hot spot, a 1 bar temperature of $166.1$\,K \citep{se98}. However, a reassessment of the Voyager radio-occultations after the corrected helium abundance as measured by the Galileo probe lead to temperatures at the occultation latitudes that are 3 to 5\,K higher \citep{gu22}, raising the possibility that the Galileo probe temperature measurement may not be strictly appropriate to the entire planet. We use a Gaussian distribution for $T_{\rm 1bar}$ centred at 165 K and with a dispersion in agreement with these observations. Measurements by the Galileo probe and recent results by the Juno mission provide constraints for the metallicity and Helium abundance in a shallow region within Jupiter's H$_2$-dominated layer. We take these constraints into account and use a He abundance of $Y_1=0.238$ \citep{vo98} and metallicity Z$_1$= 1 Z$_{\rm solar}$ in our calculations \citep{li20}. For the H$_{\rm metallic}$-dominated region, we require that the overall abundance of He is consistent with the protosolar abundance ($Y_{\rm proto}=0.277$, \citep{sa10}, and allow either different metallicity from the H$_2$-dominated region, or the same (the latter option for the dilute-core models).

\subsection{Gravity harmonics calculation}
The modelled density profiles are used to calculate the even gravity harmonics (J$_{2n}, n=1,..,5$), which are ultimately compared with the Juno-derived values to find those models consistent with the Juno observations.  The even gravity harmonics measured by Juno include contributions from both the static background state (calculated with our interior models that assume rigid body rotation) and dynamical processes (mostly from the deep atmospheric zonal winds \citep{Kaspi2017,kaspi2018,kulowski20}). We can calculate the even gravity harmonics produced from the interior density profile ($J_{\rm 2n}^{\rm static}$) by subtracting the computed dynamical contributions from the Juno measurements. 
\begin{figure}[ht]
 \begin{center}
\includegraphics[angle=0,width=0.5\textwidth]{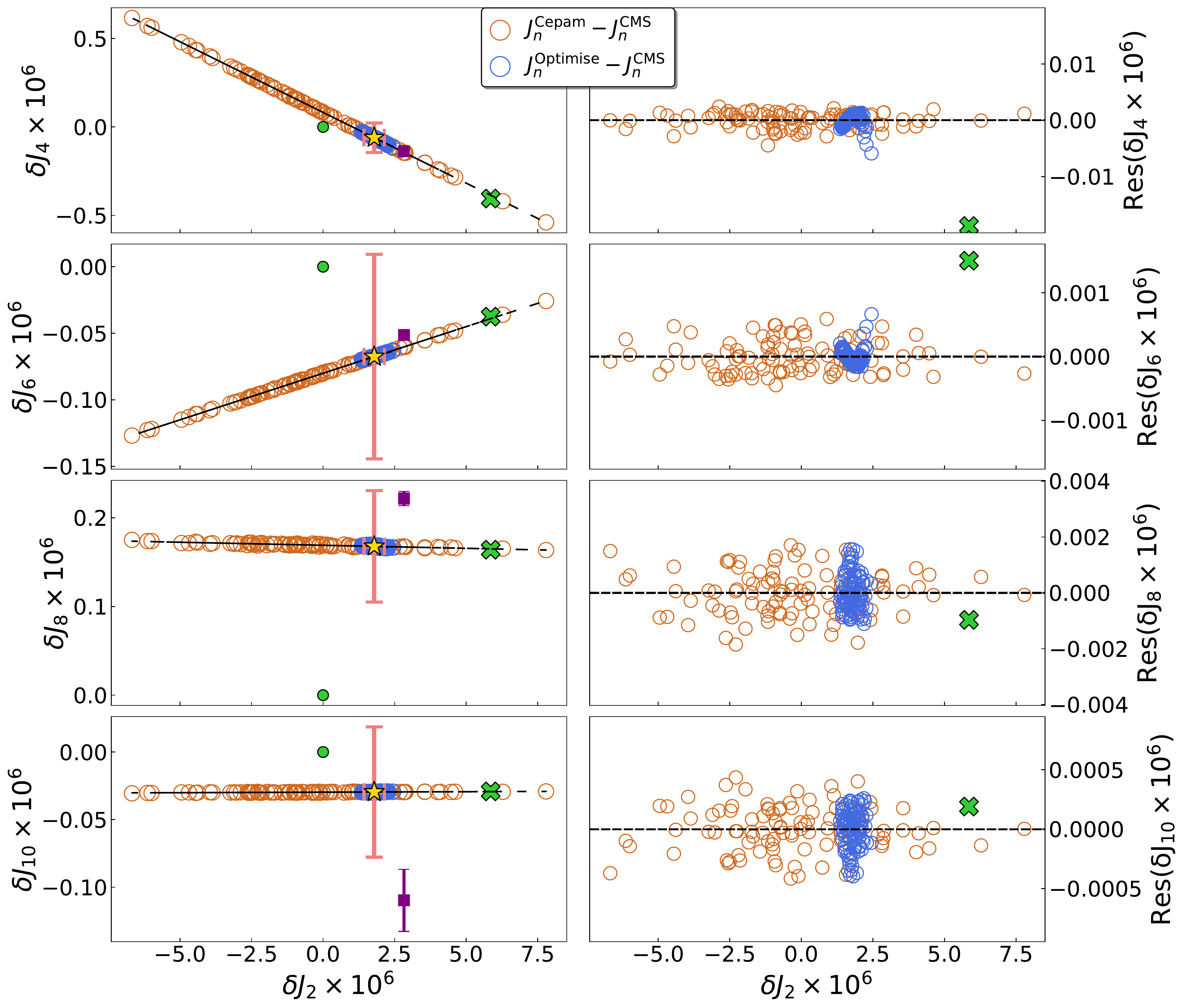}
 \end{center}
{\caption{Offsets between ToF and CMS for two subsamples (MCMC models \& optimised models). \textit{Left panels} : Offsets on the gravitational moments. The black dashed line shows the linear fit of our models. The green dot shows the origin. The yellow star corresponds to the linear fit of the optimised models median of $\Delta J_2$. The pink error-bars shows the uncertainty of the Juno measurements accounting for the differential rotation. The purple error-bars show the uncertainty of the Juno measurements. The green cross shows the former offsets from \citet{gu18}. \textit{Right panels} : Residuals.}\label{S1}}
\end{figure}
\begin{table*}\captionsetup{position=top}
\begin{center}
\caption{Offsets between ToF and CMS$^a$.}
\begin{tabular}{lccccc}
\hline & $\delta J_2 = 0$ & $\delta J_2 = \mathrm{median}(\delta J_2)$  & \citet{gu18} & Juno error & Juno w/ diff rot error\\
\hline
 $\delta J_2 \times 10^6  $ & 0 & \textbf{1.78621}   &  5.8554 & 0.014 & 0.35425\\
 $\delta J_4 \times 10^6  $ & 0.0822626 & \textbf{-0.0604954}    & -0.4045 & 0.004 & 0.0836\\
 $\delta J_6 \times 10^6  $ & -0.0799887 & \textbf{-0.0674861} & -0.0375 & 0.009 & 0.0768\\
 $\delta J_8 \times 10^6  $ & 0.16909  & \textbf{0.167862}  &  0.1641 & 0.025 & 0.0624\\
 $\delta J_{10} \times 10^6$ & -0.0297133 & \textbf{-0.0295843} & -0.0291 & 0.069 & 0.0423\\
 $\delta J_{12} \times 10^6$ & 0.004131 & \textbf{0.00411681} &  & 0.175 & \\
\hline
\end{tabular}\label{S3}
\end{center}
\footnotesize{$^a$The first column indicates the offsets for a null value of $\delta J_2$, the second (values highlighted) shows offsets estimated after picking the median value of $\delta J_2$ (the new set of offsets used in our calculations), then we show previous offsets calculated in \citet{gu18} and finally the error bars from Juno with and without accounting for differential rotation.}
\end{table*}
\begin{figure}[h]
\begin{center}
\includegraphics[angle=0,width=0.5\textwidth]{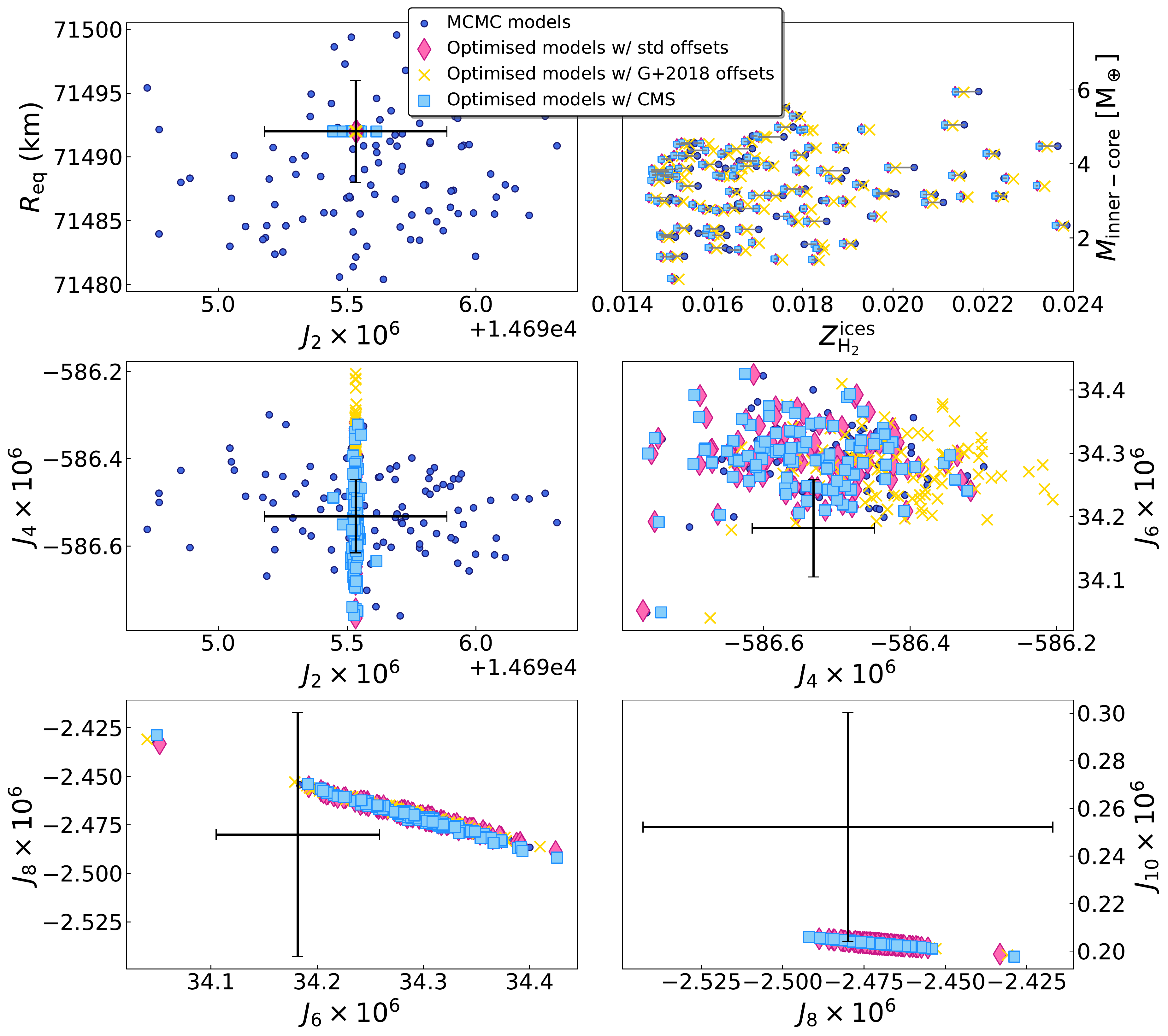}
 \end{center}
{\caption{Results for 4 sets of 100 models : random models from our preferred MCMC runs, optimised models with new offsets, optimised models with former offsets from \citet{gu18} and optimised models with CMS calculation. \textit{Top left panel} : Equatorial radius vs $J_2$. \textit{Top right panel} : Core mass vs fraction of ices in the molecular hydrogen layer. Grey lines show the pairing between each model and its optimised version. \textit{Middle and bottom panels} : Gravitational moments. The black error bars show the uncertainty of the Juno measurements accounting for the differential rotation.}\label{S2}}
\end{figure}
Therefore, the effective gravity harmonics that our interior model calculations must match are $J_{\rm 2n}^{\rm static} = J_{\rm 2n}^{\rm Juno} - \Delta J_{\rm 2n}^{\rm differential}$ ($n=1,..,5$), where $ J_{\rm 2n}^{\rm Juno}$ are the values provided by Juno mission measurements \citep{ie18,du20} and $\Delta J_{\rm 2n}^{\rm differential}$ is the contribution to the gravity harmonics due to differential rotation.  We calculate the static component of the gravity harmonics using the theory of figures (ToF) of fourth order \citep{zt78,n17} combined with an integration of the reconstructed density structure in two dimensions using a Gauss-Legendre quadrature \citep{gu18} or using the Concentric MacLaurin Spheroid (CMS) theory \citep{hu13}. Because results calculated with the CMS method are more accurate but more computationally demanding, in this paper the majority of the runs are made using the first method that allows us to perform many more runs in a shorter time, but we use the accuracy of the CMS method by calibrating the gravity harmonics obtained from the theory of figures to the CMS values \citep{gu18}. Because the gravity harmonics measured by Juno have reached a very high accuracy, estimating these offsets is essential. 

This calibration is done by assessing offsets ($\delta J_{2n} = J_{2n}(\textrm{ToF}) - J_{2n}(\textrm{CMS})$), between the gravity harmonics from the theory of figures and the ones from the CMS method, using a random sample of 100 of our preferred models. We then take these models and perform an optimisation procedure, modifying M$_{core}$ and Z$_{1}$ to perfectly fit the observational measurements of the equatorial radius $R_{eq}$ and $J_2$ -- to get offsets from the most accurate models --  and then compute the offsets for these second set of models using both methods. Figure \ref{S1} displays the offsets for both sub-samples (models with and without the optimisation procedure). Our results show a correlation between the offsets, where $\delta J_4$ and $\delta J_6$ depend strongly on $\delta J_2$. Higher order gravitational moments ($\delta J_8$, $\delta J_{10}$), are less dependent on $\delta J_2$. Thanks to these linear relationships, we can quantify the impact of an offset on $J_2$ on the offsets of the higher order gravitational moments. We also note that the residuals are very small compared to the error bars of Juno accounting for the differential rotation. We calculate the median value of $\delta J_{2}$ among the 100 optimised models and calculate the higher order offsets using the linear relationships. Table \ref{S3} lists the offsets found, where we note that previous offsets from \citet{gu18} almost lie on the linear regression curves, showing that our offsets have changed very slightly and thus our calculations are robust.

\subsection{Dynamical contribution to the gravity harmonics}\label{section:diffrot}
In order to provide the plausible range of dynamical contribution to the even gravity harmonics ($\Delta J_{\rm n}^{\rm differential}$), we consider the widest possible range of flow profiles that still match the gravity harmonics. The Juno gravity measurements revealed significant north-south asymmetries in Jupiter\textquoteright s gravity field \citep{ie18}. Such hemispherical asymmetries (measurable $J_{3}$, $J_{5}$, $J_{7}$ and $J_{9}$) can only be due to interior flows. The resemblance of the measured gravity signature of the flow to that obtained by extending the east-west (zonal) observed cloud-level winds inward \citep{Kaspi2013}, suggested that not only the flows reach deep ($\sim3000$~km which are $\sim100$~kbar), but that the flow pattern is generally similar to the flow profile at the cloud level \citep{kaspi2018,kaspi2020}. Although variations from this profile at depth are possible, it is statistically unlikely that the deep flow profile varies significantly and will still match all four measured odd gravity harmonics \citep{Tollefson2017,Duer2020}. Any deep flow profile will also have a signature in the even gravity harmonics, providing a dynamical contribution to the measurements beyond that coming from the interior density profile. 

\begin{figure}
  \begin{center}
\includegraphics[angle=0,width=0.5 \textwidth]{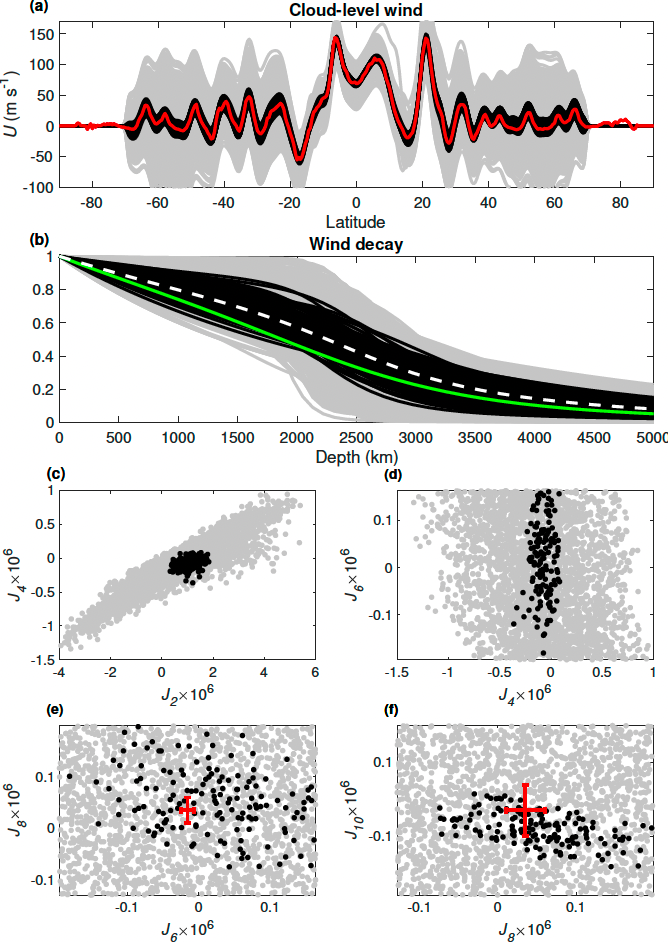}
 \end{center}
  \caption{Contribution to the gravity harmonics from the deep atmospheric flow. Shown are the solutions for the cloud-level wind (a), the radial decay profile (b), and the wind-induced even gravity harmonics (c-f). The grey lines/dots show all of the 3000 cases, and the black lines/dots show the plausible solutions. The red line in (a) shows the observed cloud-level wind, and the green line in (b) shows the solution in \cite{kaspi2018}. The red crosses in (e-f) show the expected mean value of the differential contribution.}
 \label{fig:diffrot}
\end{figure}

To determine such dynamical contribution to the even gravity harmonics we retrieve the wind field from the gravity field data using an adjoint inverse technique \citep{Galanti2016,Galanti2017b}. We allow the zonal wind structure to vary both in its latitudinal profile (from the observed cloud-level wind structure) and its vertical profile \citep{Galanti2019a}. This is achieved through randomly varying $\Delta J_{\rm 2n}^{\rm differential}, n=1,5$ within a range of 30$\sigma$ of the Juno values (we choose a large range because it is not known what fraction of the even harmonic measurements is coming from the dynamics \citet{Kaspi2017}). Every profile is extended inward along the direction of the spin axis, assuming a radial decay profile, and then given that the flow must be in thermal wind balance with the density field, the balancing density field is integrated to give the dynamical gravity harmonics \citep{Kaspi2016}. For each of the 3000 cases, we calculate the best fitting wind profile that is also consistent with the odd gravity harmonics, thus each matching seven gravity values (odd harmonics $J_{3}^{\rm Juno}$ to $J_{9}^{\rm Juno}$, and even differential contributions  $\Delta J_2^{\rm differential}$ to $\Delta J_{10}^{\rm differential}$). Once all solutions are obtained (gray profiles in Fig. \ref{fig:diffrot}a,b, for the meridional and vertical profiles, respectively), we select as plausible solutions those where the meridional profile of the zonal wind is within 20~m s$^{-1}$ of the observed cloud-level profile (black profiles in Fig. \ref{fig:diffrot}a,b). This uncertainty is consistent with the measurement error \citep{kaspi2020}. The resulting range of physically plausible dynamical contribution to the even gravity harmonics is: $\Delta J_{2}^{\rm differential} = 1.039\pm0.354$, $\Delta J_{4}^{\rm differential} =-0.076\pm0.083$, $\Delta J_{6}^{\rm differential} = 0.016\pm0.076$, $\Delta J_{8}^{\rm differential} =0.053\pm0.062$, $\Delta J_{10}^{\rm differential} =-0.080\pm0.042$, (where $\Delta J_{2n}^{\rm differential}/10^{-6}$).

\section{Results}
\subsection{Envelope inhomogeneity}
\begin{figure*}
 \begin{center}
\includegraphics[angle=0,width=0.65\textwidth]{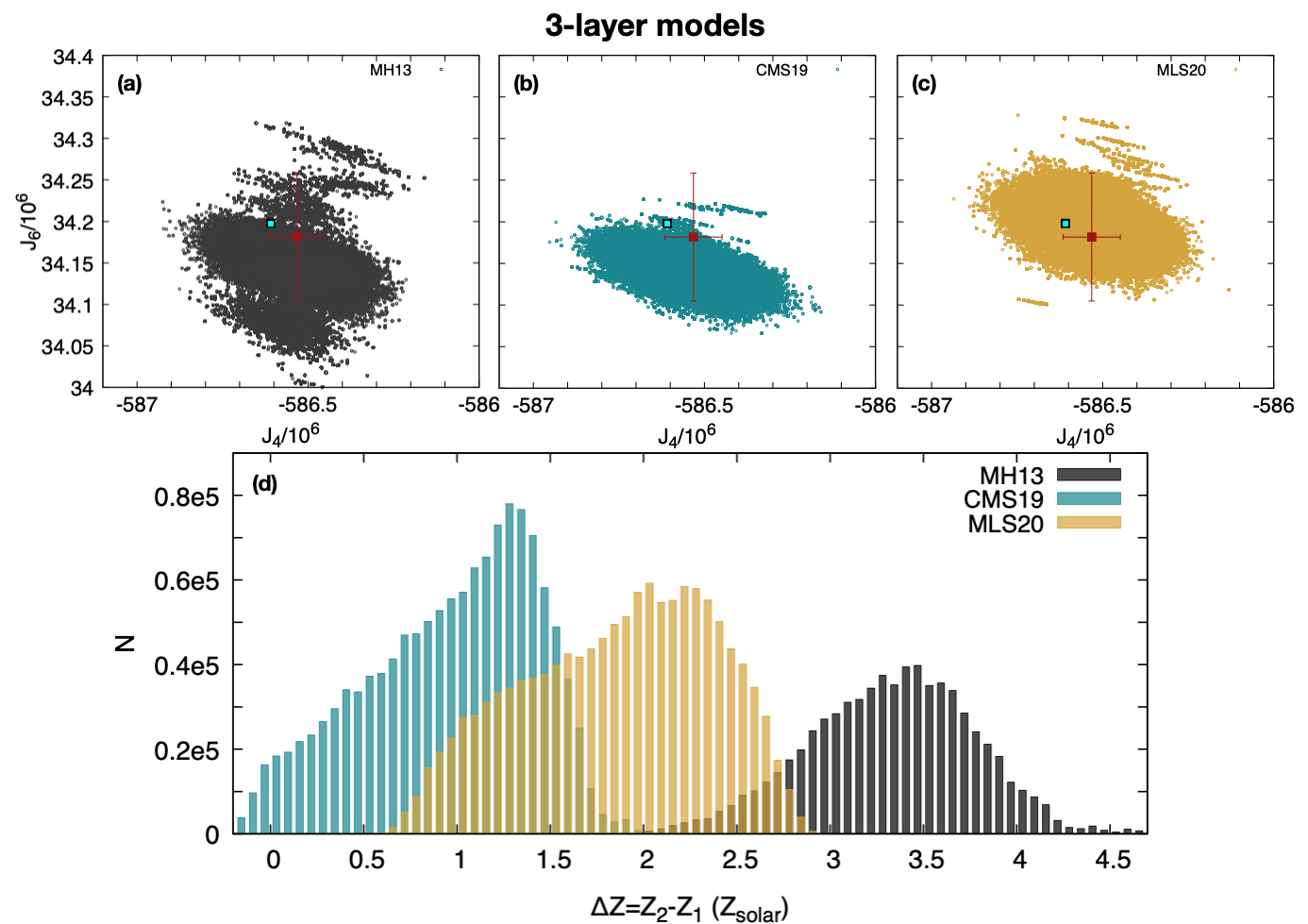}
 \end{center}
{\caption{Details of models with 3 layers. Panels (a), (b) and (c): J$_4$ vs. J$_6$ for 3-layer models calculated using different equations of state for H, indicated in the Figure. Cyan squares show the Juno measurements ($J_{\rm 2n}^{\rm Juno}$), and gravitational harmonics corrected by differential rotation --the ones to match with our interior models ($J_{\rm 2n}^{\rm static}$)-- are shown in red. Panel (d): Histogram showing the difference in heavy elements ($\Delta $Z$ = $Z$_2 - $Z$_1$) between the H$_{\rm metallic}$- and the H$_2$-dominated layers.}\label{Fig1:3L}}
\end{figure*}
\begin{figure*}
 \begin{center}
\includegraphics[angle=0,width=0.65\textwidth]{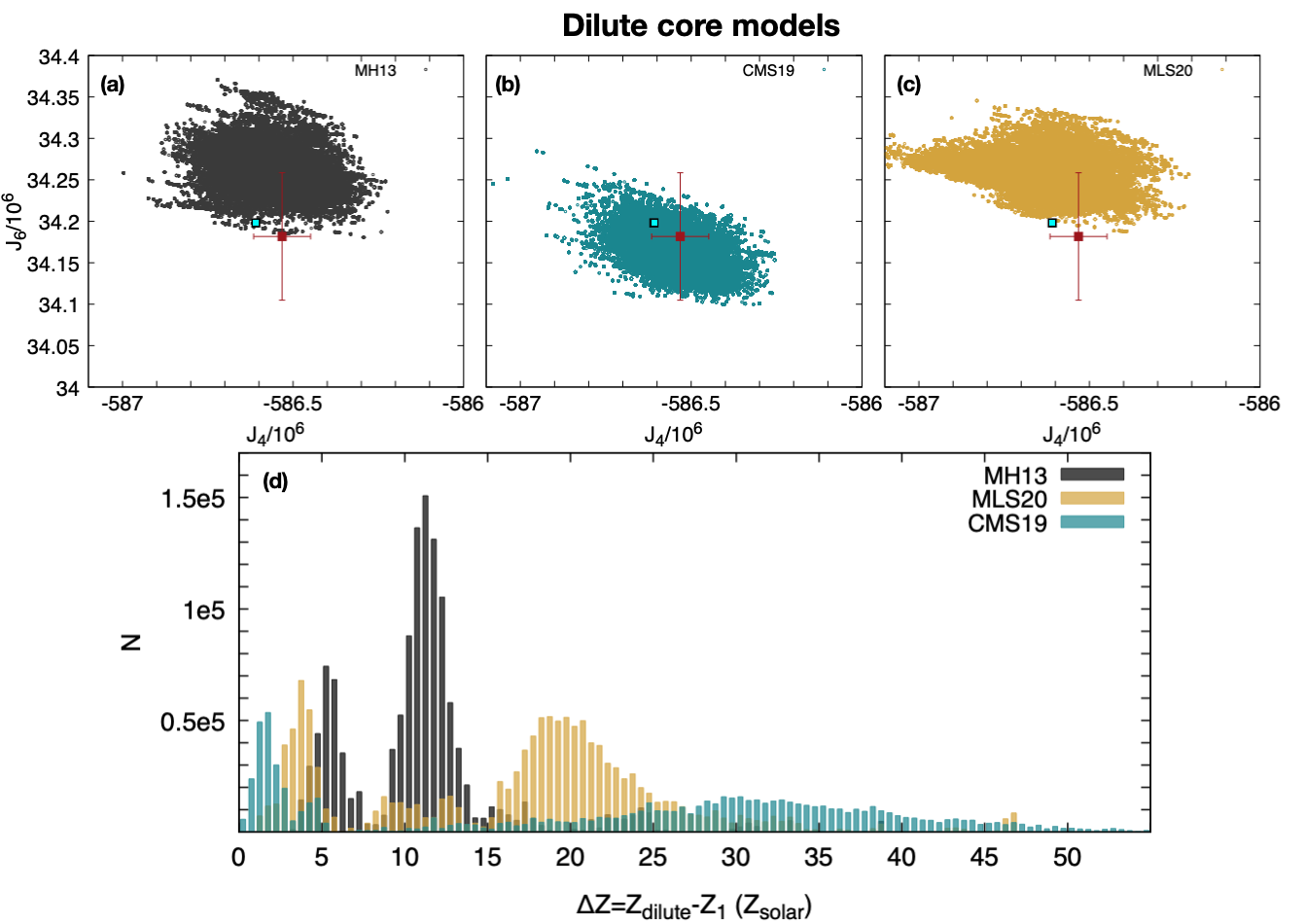}
 \end{center}
{\caption{Specifics of models with dilute core. Panels (a), (b) and (c): J$_4$ vs. J$_6$ obtained considering different equations of state for H. The Juno measurement (J$_{\rm 2n}^{\rm Juno}$) is indicated with the cyan square. The effective gravity harmonics to be matched by our interior models (J$_{\rm 2n}^{\rm static}$) are shown in red. Panel (d): Histogram that shows the difference in heavy elements ($\Delta $Z$ = $Z$_{\rm dilute} - $Z$_1$) between the dilute core region and the H$_2$-dominated layer.}\label{Fig2:Dilute}}
\end{figure*}

All of our models reproduce $J_2$ and Jupiter's radius, and Figures \ref{Fig1:3L} (panels a,b,c) and \ref{Fig2:Dilute} (panels a,b,c) show the values of $J_4$ and $J_6$ for the MCMC solutions in the case of a 3-layer model and with a dilute core model, respectively (the other $J_{\rm 2n}$ are shown in Methods). We also note that all our models are consistent with metallicities observed by Juno and the helium abundance in Jupiter's atmosphere as measured by the Galileo probe. When analysing the distribution of heavy elements in the envelope we consider that the envelope consists of the H$_2$- and the  H$_{\rm metallic}$-dominated layers for the 3-layer models, and it consists of the H$_2$-, the H$_{\rm metallic}$-dominated layers and the dilute core, for the dilute core models. In Figure \ref{Fig1:3L}d, we show results of 3 layer models where we see that the difference in heavy elements between the H$_2$- and the  H$_{\rm metallic}$-dominated layers ($\Delta Z = Z_2 - Z_1$) is larger than zero for all models calculated with the MH13 and MLS20 equations of state. For the CMS19 equation of state, we find that the probability of finding models with $\Delta Z>0$ is of 97.6$\%$. 

Figure \ref{Fig2:Dilute}d shows the difference in heavy elements within the envelope for models with a dilute core. In this case, we calculate the difference in heavy elements between the H$_2$- and the dilute core region ($\Delta Z = Z_{\rm dilute} - Z_1$). We see that all models have $\Delta Z>0$, independently of the equation of state used in the calculations. Our results robustly demonstrate that Jupiter's envelope is not homogeneous: the external layer of the envelope is depleted of heavy elements compared to the inner parts of the envelope. This result is independent on the models adopted for the interior of the planet and the equation of state used. We note, however, that different equations of state lead to different distributions of the heavy elements in the interior of the planet. 

\begin{figure*}
 \begin{center}
\includegraphics[angle=0,width=0.95\textwidth]{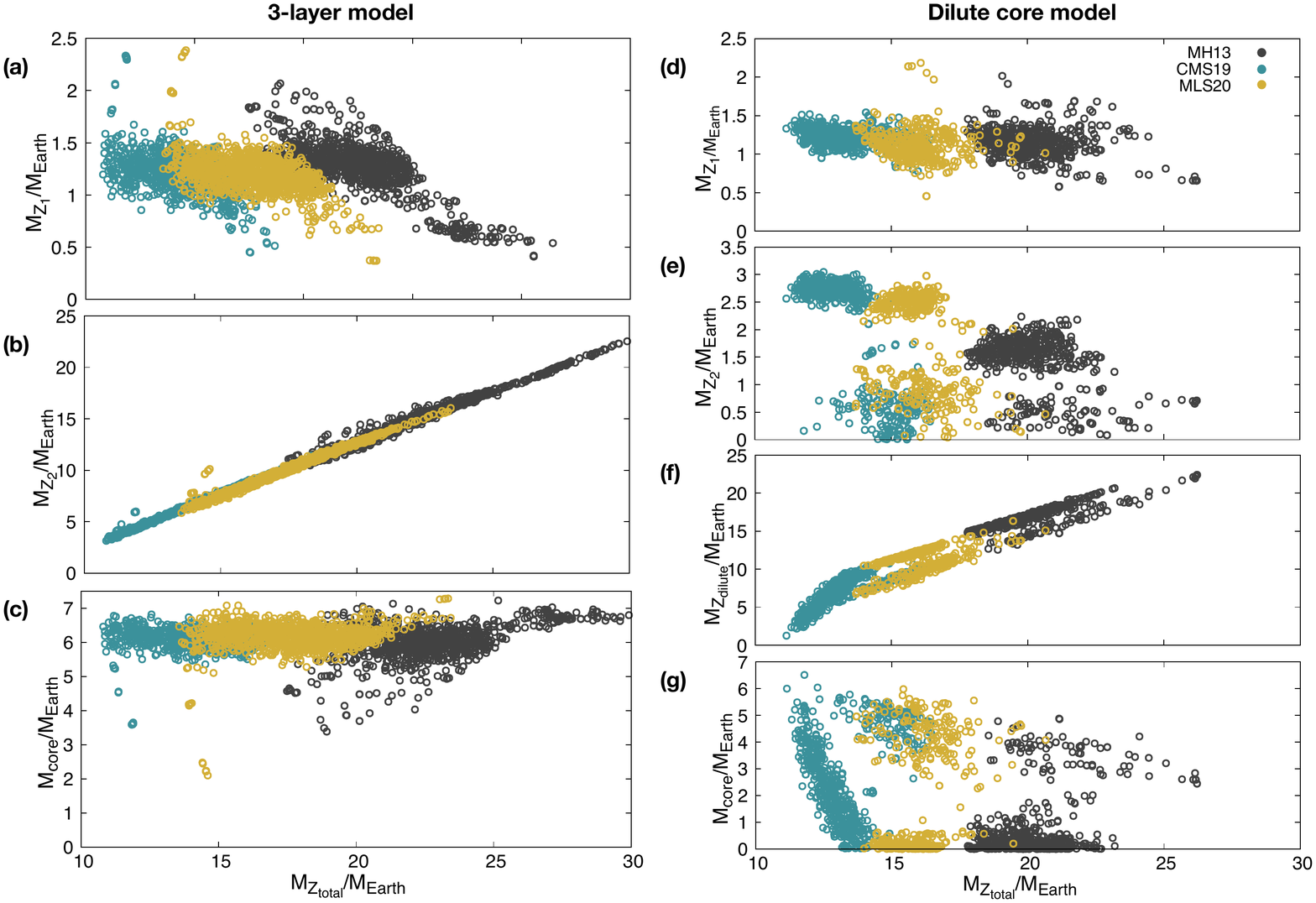} 
  \end{center}
{\caption{Mass of heavy elements in the different layers as a function of the total mass of heavy elements in Jupiter for a random sample of 1000 models extracted from our 3 layer models (a, b and c) and from models with a dilute core (d, e, f, and g). Different colours show models calculated with different equations of state. Panels (a) and (d) show the mass in the H$_2$-dominated region in the y-axis, panels (b) and (e) show the mass in the H$_{\rm metallic}$-dominated region, panel (f) shows the mass in the dilute core and panels (c) and (g) show the mass in the inner core.}\label{Fig3:Masses}}
\end{figure*}

\subsection{Distribution and mass of the heavy elements}
An analysis of the mass of heavy elements in the different layers for a random sample of 1000 of our models is shown in Figure \ref{Fig3:Masses}. We see that the total mass of heavy elements (M$_{\rm Z_{\rm total}}$) varies between 11 and 30 M$_{\rm Earth}$, with differences resulting from the choice of the equation of state. Calculations done with MH13 equation of state have $18<$M$_{\rm Z_{\rm total}}<30 $M$_{\rm Earth}$, models with MLS21 have $14<$M$_{\rm Z_{\rm total}}<24 $M$_{\rm Earth}$ and models with CMS19 have $11<$M$_{\rm Z_{\rm total}}<18 $M$_{\rm Earth}$, independently of the model of Jupiter adopted (3-layers or dilute core). For the 3-layer models, the differences mostly arise from discrepancies of the mass of heavy elements in the H$_{\rm metallic}$-dominated region (M$_{\rm Z_2}$) that varies between 2 and 23 M$_{\rm Earth}$, depending on the equation of state. For models with a dilute core, the differences are mostly due to differences in the mass of heavy elements in the dilute core region (M$_{\rm Z_{\rm dilute}}$), that is found to vary between 1 and 25 M$_{\rm Earth}$, depending on the equation of state adopted in the calculation. The mass of heavy elements in the  H$_2$-dominated region (M$_{\rm Z_1}$) is quite similar for models with different equations of state, independently on the model adopted for Jupiter. This is expected given the prior used to match the observational constraints from the Juno and Galileo missions. Regarding the inner core, we find that it varies between 0 and 7  M$_{\rm Earth}$ for all models, with its exact value depending on the model adopted for Jupiter's interior. The 3-layer models have M$_{\rm core}\simeq 6\ $M$_{\rm Earth}$, independently on the equation of state adopted. Conversely, we see two groups of solutions for models with a dilute core: a group with inner core masses between 3 and 6 M$_{\rm Earth}$, and another group with small masses up to 2 M$_{\rm Earth}$. 

Interestingly, models tend to have temperatures at 1 bar going from $\sim$169 to $\sim$188 K for the 3-layer models to values between $\sim$165 and $\sim$182 K for the dilute core models, depending on the equation of state adopted (see Sect. \ref{Appendix}). While these values are high compared to observations by the Galileo probe ($166.1$\,K, \citet{se98}), it is not clear whether those in situ measurements represent the typical 1 bar temperature on Jupiter because of latitudinal variability, which may exist to a limited extent \citep{Fletcher+2020}. Furthermore, a reassessment of the Galileo probe data lead to an increase of the temperature of $\simeq$4 K \citep{gu22} and, additionally, the possibility of superadiabaticity in the interior \citep{Leconte+2017,Guillot+2020b} could yield a deep entropy corresponding to a temperature a few degrees higher than the measured value at 1 bar. In all cases dilute core models have temperatures more in agreement with the expectations given the values observed and the uncertainties in this parameter. Regarding the separation pressure for the immiscibility of helium in hydrogen, our values are always close to 3 Mbar (see Appendix), that are in agreement with the higher limit of numerical calculations \citep{mo13,scr18} and more in agreement with recent laboratory experiments estimations \citep{br21}.

\subsection{Comparison with previous models}
In this paper we performed a large exploration of the parameter space. We find that our results include most other individual models presented in other works.

Dilute core models were first modelled by \citet{wa17}, who, using the MH13 EoS, found a value of M$_{\rm core}+M_{\rm dilute}$ between 10 and 24 M$_{Earth}$ and M$_{\rm Z_{total}}$ of 24-27 M$_{Earth}$. These results are consistent with our results, even though most of them have sub-solar atmospheric metallicities.

\citet{Ni19} performed 4-layer models using MH13 and CMS19 EoS. They found a range of M$_{\rm core}$+M$_{\rm dilcore}$ of 6.5-27 M$_{Earth}$ and M$_{\rm Z_{total}}$ of 24-28 M$_{Earth}$ with MH13 EoS and M$_{\rm core}$+M$_{\rm dilcore}$ of 3-12 M$_{Earth}$ and M$_{\rm Z_{total}}$ of 8-12 M$_{Earth}$ with CMS19 EoS. For both EOS, the results are in good qualitative agreement with ours, even though, in this case again, their optimisation led to sub-solar atmospheric metallicities as opposed to our results where $Z_{1} = 1 \ Z_{\rm solar}$.

\citet{de19} imposed the constraint of a minimum atmospheric metallicity $Z=0.02$. They found that in order to fit this and all other constraints using the CMS19 EOS, an inward-decrease of the abundance of heavy elements between the H$_{2}$- and H$_{\rm metallic}$-dominated regions was required. We also find these type of solutions when using CMS19 EoS and 3-layer models (Figure \ref{Fig1:3L}), but since they represent $2.4\%$ of our sample we find this unlikely. The M$_{\rm Z_{total}}$ found by \citet{de19} are of 25-30 and 40-45 M$_{Earth}$ for models without and with inner core, respectively. Our results with CMS19 always lead to considerably smaller values. A similar discrepancy with the \citet{de19} study is found by \cite{ne21}.

\citet{ne21} used CMS19 and dilute core models. They found M$_{\rm core}$ of up to 3.8 M$_{Earth}$ and M$_{\rm Z_{total}}$ up to 13 M$_{Earth}$, comparable to our results. We note that in \citet{ne21} the authors calculate the gravitational harmonics using an expansion of the ToF of seventh order, different from the method used here (section \ref{section:diffrot}). Similarly to our conclusions, they also find that higher temperatures at 1 bar help in reaching higher metallicities in the atmosphere. Their models have separation pressures between the H$_{2}$- and H$_{\rm metallic}$-dominated regions close to 6 Mbar while our models use separation pressures around 3 Mbar.

\section{Conclusions}
This study comprehensively reproduces observational constraints from the Juno measurements (the even and odd gravity harmonics and water abundance in the atmosphere), along with helium measurements from the Galileo probe, exploring different models for Jupiter's interior and considering all recent equations of state. We show that the gravity constraints point to a deep entropy of Jupiter that corresponds to a 1 bar temperature that is higher than traditionally assumed, i.e. 170 to 180\,K rather than 166\,K. We robustly demonstrate that the heavy element abundance is not homogeneous in Jupiter's envelope. Our results imply that Jupiter continued to accrete heavy elements in large amounts while its hydrogen-helium envelope was growing, contrary to predictions based on the pebble-isolation mass in its simplest incarnation \citep{la14}, favouring instead planetesimal-based or more complex hybrid models \citep{al18,octi20}. Furthermore, the envelope did not mix completely during the planet's subsequent evolution, not even when Jupiter was young and hot \citep{va18}. Our result clearly shows the need for further exploration of non-adiabatic interior models for the giant planets and it provides a base example for exoplanets: a non-homogeneous envelope implies that the metallicity observed is a lower limit to the planet bulk metallicity. Therefore, metallicities inferred from remote atmospheric observations in exoplanets might not represent the bulk metallicity of the planet. Moreover, we demonstrate that knowledge of the equation of state is crucial in determining the mass of heavy elements in the interior of Jupiter and we put important constraints on Jupiter's inner core, which is found to be up to 7 M$_{\rm Earth}$, a result that is independent on the interior model and equation of state adopted in the calculations.

\section*{Acknowledgments}
JIL was supported by the Juno Project through a subcontract with the Southwest Research Institute.

\appendix \label{Appendix}
\section{Bayesian Statistical Model}
Taking into account the uncertainties in the equation of state, the temperature at 1 bar, the pressure where helium-rain happens and the error bars on the measurements themselves, we employ a Bayesian statistical approach to explore a large ensemble of Jupiter models and find which are the possible scenarios for Jupiter's internal structure that fit all observational constraints. The parameters of CEPAM we allow to vary are:
\begin{equation}
\boldsymbol{\theta} = \begin{cases}
[M_{core}, Z_{2}^{\rm rock}, Z_{2}^{\rm ice}, T_{\rm 1bar}, P_{\rm He}, \Delta T_{\rm He}] & \text{3-layer}\\
[M_{core}, Z_{\rm dilute}^{\rm rock}, Z_{\rm dilute}^{\rm ice}, m_{\rm dilute}, T_{\rm 1bar}, P_{\rm He}, \Delta T_{\rm He}] & \text{dilute-core}
\end{cases}
\end{equation}
and the data 
\begin{equation}
\boldsymbol{X}= [ J_{2}^{static}, J_4^{static}, J_6^{static}, J_8^{static}, J_{10}^{static}, R_{eq}^{obs}]
\end{equation}

The quantity of interest is the posterior density function (PDF) of the parameters conditional on the data, $p(\boldsymbol{\theta}|\boldsymbol{X})$. Bayes' theorem tells us that it is completely specified by the likelihood, $p(\boldsymbol{X}|\boldsymbol{\theta})$, and the prior density function, $p(\boldsymbol{\theta})$, through the relation
\begin{equation}
  \displaystyle
  p(\boldsymbol{\theta}|\boldsymbol{X}) \propto p(\theta)p(\boldsymbol{X}|\boldsymbol{\theta}).
\end{equation}
The likelihood is the probability density of the data, conditional on the parameters. In general, it is not a probability density for the parameter. The prior density encodes the information we have on the parameters before the inference processes (e.g. theoretical limits, previous measurements).

It is possible, under some assumptions, to write down analytical forms for the likelihood and the prior. To define the former, we first assume an additive statistical model
\begin{equation}
  \displaystyle
  \boldsymbol{X} = \boldsymbol{\jmath}(\boldsymbol{\theta}) + \boldsymbol{\epsilon}
\end{equation}
Here $\boldsymbol{\jmath}$ is a mapping from the space of parameters to the space of observables. In practice, it will represent our code solving the equations of planetary structure. The vector $\boldsymbol{\epsilon}$ represent a random noise. The distribution of this noise is thus the distribution of $\boldsymbol{X} - \boldsymbol{\jmath}(\boldsymbol{\theta})$. We assume now that the distribution of $\boldsymbol{\epsilon}$ is normal and that its covariance matrix is diagonal. The likelihood function is therefore
\begin{equation}
  \displaystyle
  p(\boldsymbol{X}|\boldsymbol{\theta}) \propto \prod_i \exp{\left[ -\frac{(X_i - \jmath_i(\boldsymbol{\theta}))^2}{2\sigma^2_i} \right] }. 
\end{equation}
The $\sigma_i^2$ represent the diagonal elements of the covariance matrix and are given by the observational uncertainties corresponding to the components of $\boldsymbol{X}$. The adopted values are given in Tables S1 and S2. 

In order to simplify the prior density, we assume that all parameters are independent. It can thus be written as the product of univariate densities
\begin{equation}
  \displaystyle
  p(\boldsymbol{\theta}) \propto \prod_i p_i(\theta_i). 
\end{equation}
The chosen priors are given in Tables \ref{tableS1} and \ref{tableS2}. The boundaries of all priors were chosen using a test-and-trial stage, ensuring that they were wide enough to avoid numerical issues during the sampling stage and that they were narrow enough so that the parameter space to sample is not too vast. 

\begin{table*}
\centering
\caption{Parameters explored in our mcmc calculations for 3-layer models. The parameter is given in the first column, the corresponding distribution in the second, the lower and upper bounds in the third and fourth. When relevant, the mean and the standard deviation of the truncated normal are given in columns five and six. Y$_{\rm proto}$=0.277, Y$_1$=0.238, Z$_1$=0.0153.}
\begin{tabular}{@{}lcccccc@{}}
\hline
Parameter &  Distribution & Lower bound & Upper bound & $\mu$ & $\sigma$\\
\hline
\hline
$M_{core}$ (M$_{Jup}$)  &Uniform & 0 & 0.075     & -- & -- \\
$Z_{2}^{\rm rock}$ &Uniform & 0 & 0.5     & -- & -- \\
$Z_{2}^{\rm ice}$  &Uniform & 0 & 0.5     & -- & -- \\
$T_{\rm 1bar}$ (K) &Normal  & 135 & 215    & 165 & 4 \\
$P_{\rm He}$ (Mbar) &Normal & 0.8 & 9 & 3 & 0.5 \\
$\Delta T_{\rm He}$ (K) & Uniform& 0 & 2000 & -- & -- \\
\hline
\end{tabular} \label{tableS1}
\begin{flushleft}
\end{flushleft}
\end{table*}
\begin{table*}
\centering
\caption{Parameters explored in our mcmc calculations for dilute core models. The parameter is given in the first column, the corresponding distribution in the second, the lower and upper bounds in the third and fourth. When relevant, the mean and the standard deviation of the truncated normal are given in columns five and six. $Y_{proto}=0.277$, $Y_{H_2}=0.238$, $Z_{1}=0.0153$, $Z_{2}=0.0153$}.
\begin{tabular}{@{}lcccccc@{}}
\hline
Parameter &  Distribution & Lower bound & Upper bound & $\mu$ & $\sigma$\\
\hline
\hline
$M_{inner-core}$ (M$_{Jup}$)  &Uniform & 0 & 0.075     & -- & -- \\
$Z_{\rm dilute}^{\rm rock}$ &Uniform & 0 & 0.5     & -- & -- \\
$Z_{\rm dilute}^{\rm ice}$  &Uniform & 0 & 0.5     & -- & -- \\
$m_{\rm dilute}$  &Uniform & 0 & 0.6     & -- & -- \\
$T_{\rm 1bar}$ (K) &Normal  & 135 & 215    & 165 & 4 \\
$P_{\rm He}$ (Mbar) &Normal & 0.8 & 9 & 3 & 0.5 \\
$\Delta T_{\rm He}$ (K) & Uniform& 0 & 2000 & -- & -- \\
\hline
\end{tabular}\label{tableS2}
\begin{flushleft}
\end{flushleft}
\end{table*}

We used an affine invariant Markov chain Monte Carlo algorithm \citep{Christen2010,Goodman2010} in order to sample the space of parameter and approximate the posterior probability functions. It uses the relative positions in the parameter space of several Markov chains, run in parallel, to efficiently adapt the proposition law. Using 512 walkers, we found that good convergence for the MCMC simulations can be reached with approximately 10000 iterations of the algorithm.

\begin{figure*}
  \centering
\includegraphics[angle=0,width=1 \textwidth]{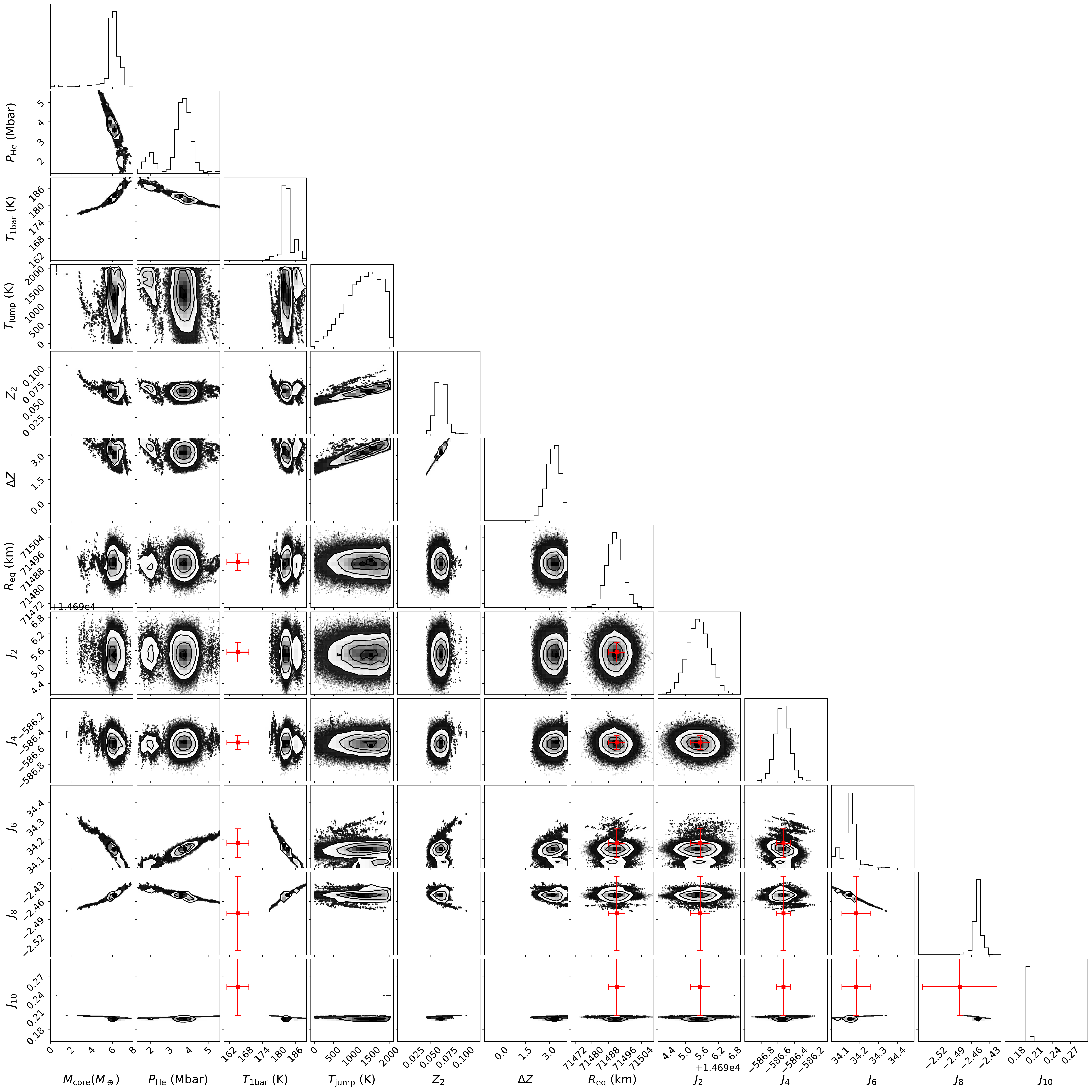}
  \caption{MCMC corner plot showing the posterior distribution of variables obtained with the 3-layer model and the MH13 equation of state. Red points with error bars show the observed parameters as a reference.}
  \label{S4}
\end{figure*}

\begin{figure*}
\centering
\includegraphics[angle=0,width=1 \textwidth]{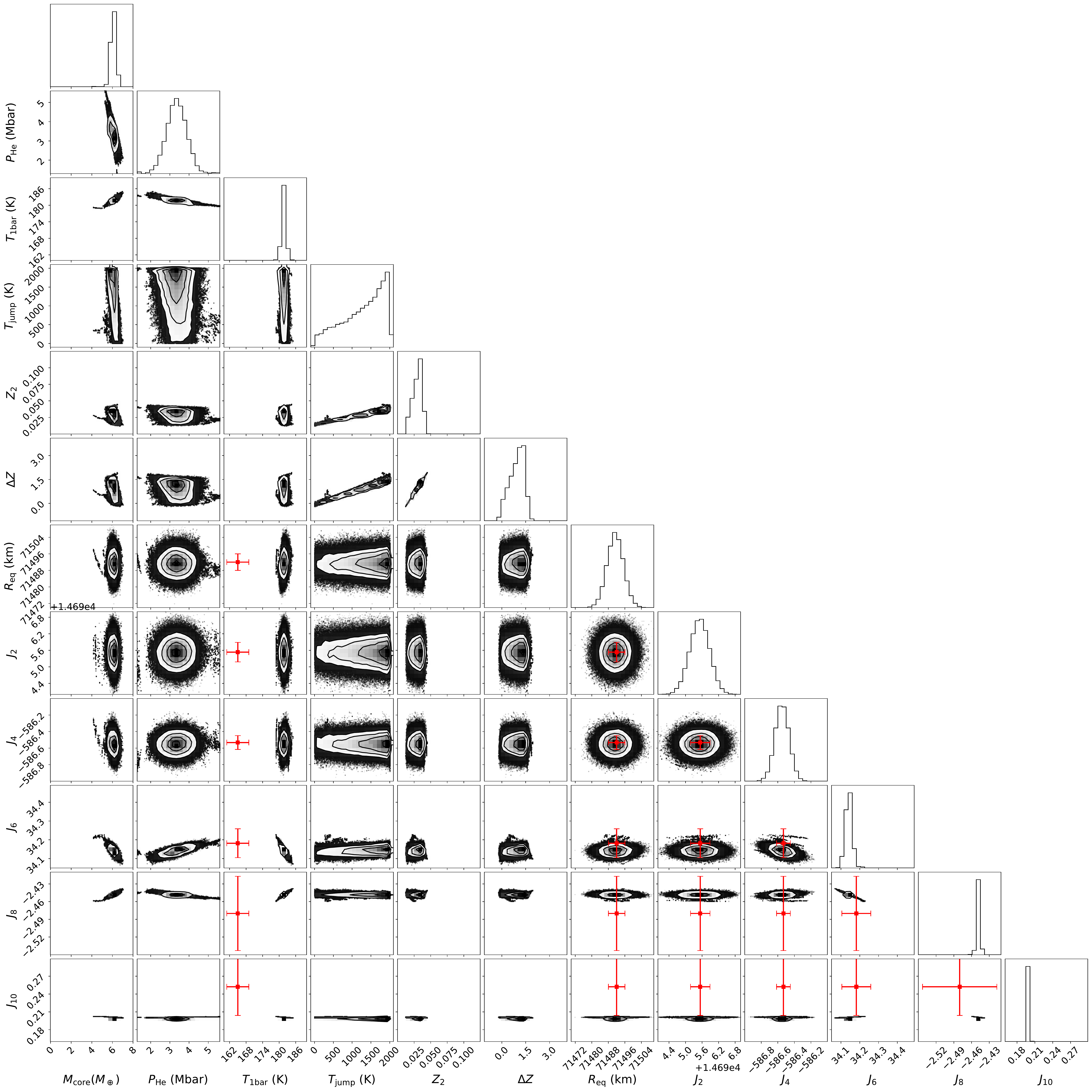}
  \caption{Posterior distribution resulting of the MCMC runs with 3-layers and CMS19 equation of state. Observed parameters are indicated with the red points.}
  \label{S5}
\end{figure*}

\begin{figure*}
\centering
\includegraphics[angle=0,width=1 \textwidth]{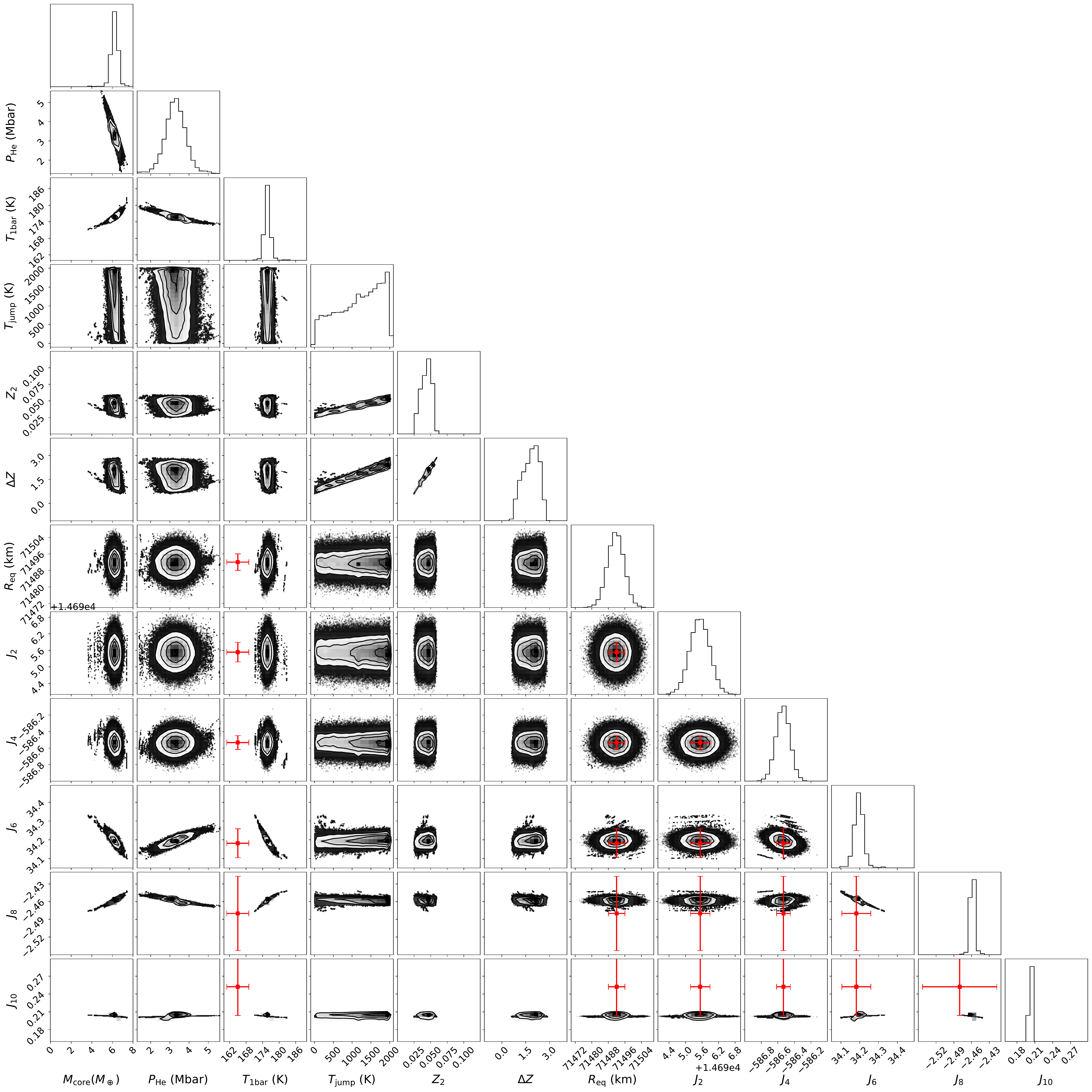}
  \caption{Distributions resulting of the MCMC runs with 3-layers and MLS20 equation of state.}
  \label{S6}
\end{figure*}

\begin{figure*}
\centering
\includegraphics[angle=0,width=1 \textwidth]{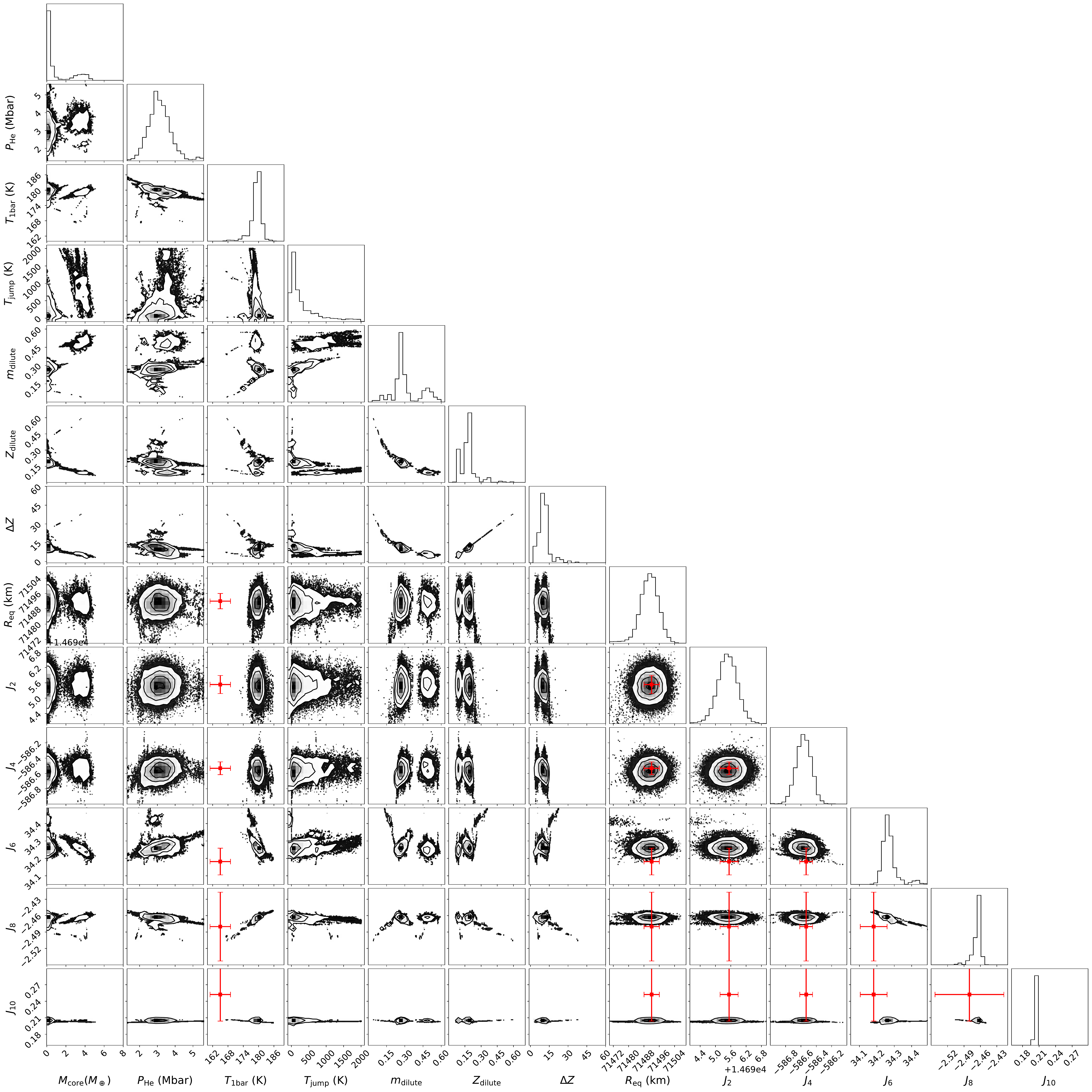}
  \caption{Corner plot resulting from the runs with the dilute-core model and MH13 equation of state. Red dots show observed parameters.}
  \label{S7}
\end{figure*}

\begin{figure*}
\centering
\includegraphics[angle=0,width=1 \textwidth]{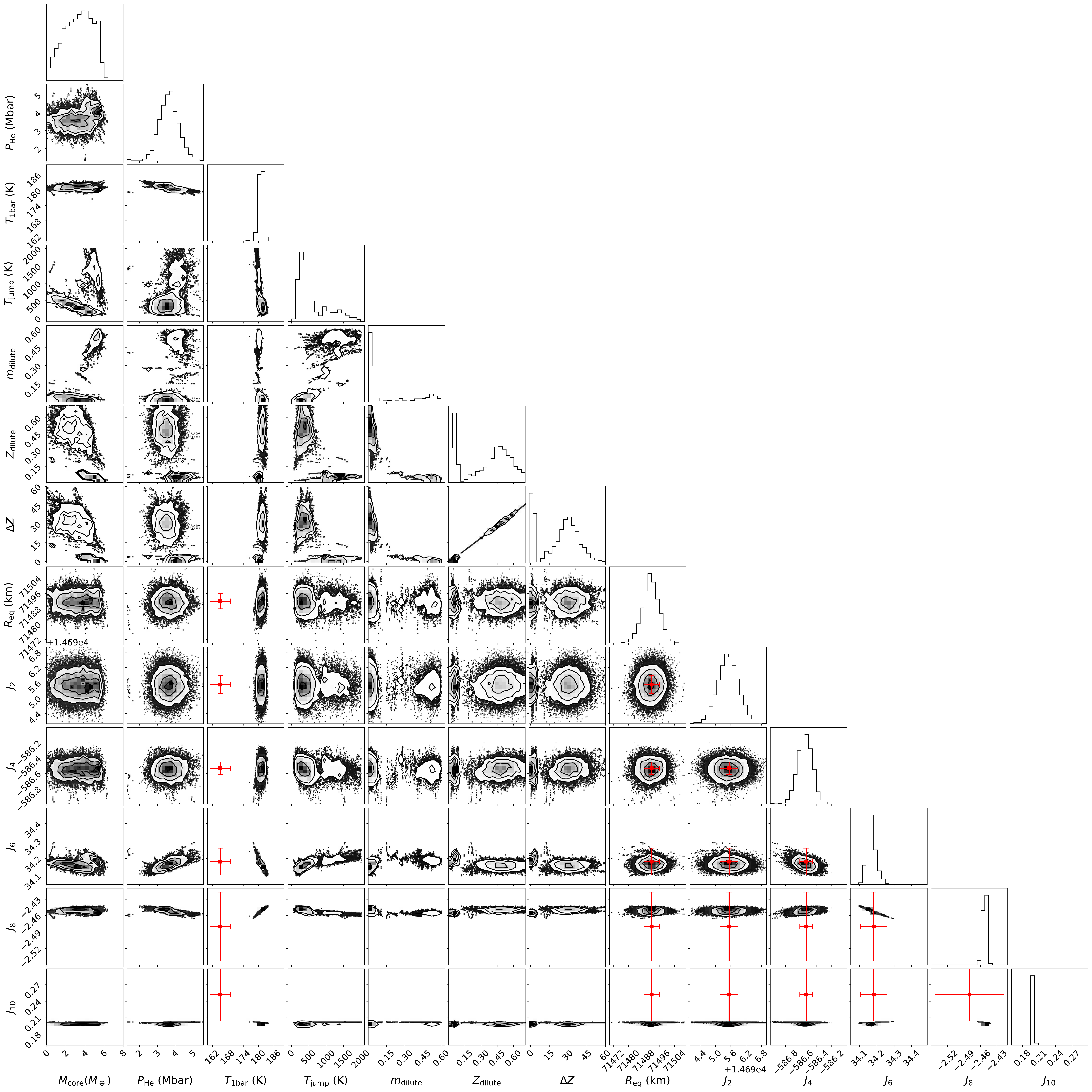}
  \caption{Distribution of different parameters resulting from the runs with the dilute-core model and CMS19 equation of state.}
  \label{S8}
\end{figure*}

\begin{figure*}
\centering
\includegraphics[angle=0,width=1 \textwidth]{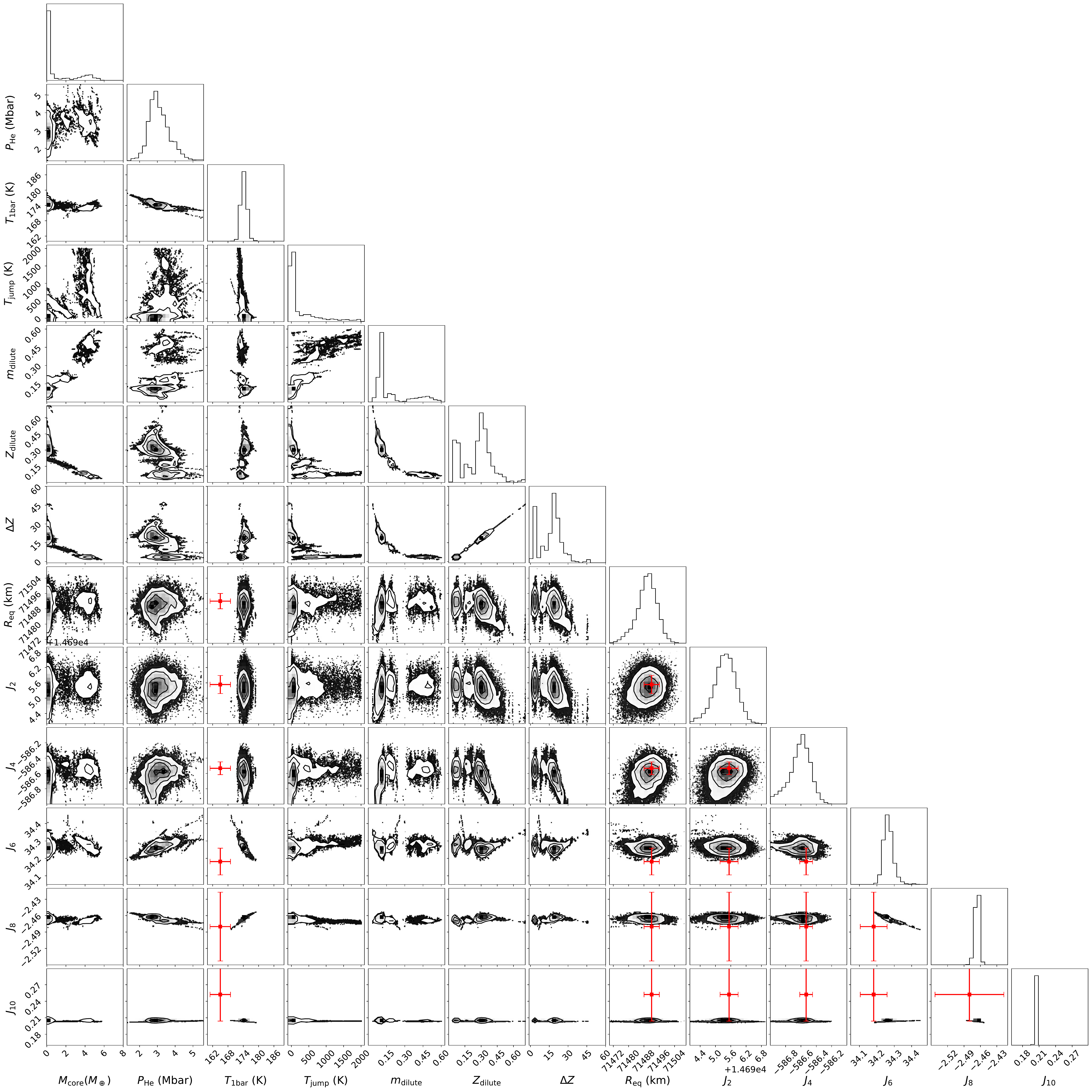}
  \caption{Resulting distributions obtained with the dilute-core model and MLS20 equation of state.}
  \label{S9}
\end{figure*}

Figures \ref{S4}, \ref{S5}, \ref{S6}, \ref{S7}, \ref{S8} and \ref{S9} show the corner plots with all parameters that result of the MCMC calculations for the six cases considered (see main text for an extended discussion on the results). 

\begin{thebibliography}{}
\bibitem[Alibert et al.(2018)]{al18} Alibert, Y., Venturini, J., Helled, R., et al.\ 2018, Nature Astronomy, 2, 873. 
\bibitem[Bazot et al.(2012)]{ba12}Bazot, M., Bourguignon, S., \& Christensen-Dalsgaard, J.\ 2012, \textit{MNRAS}, 427, 1847.
\bibitem[Brygoo et al.(2021)]{br21}Brygoo, S., Loubeyre, P., Millot, M. et al.\ 2021, Nature, 593, 517–521.
\bibitem[Chabrier et al.(2019)]{cms19}{Chabrier}, G. and {Mazevet}, S. and {Soubiran}, F.\ 2019, ApJ, 872, 1.
\bibitem[Christen \& Fox(2010)]{Christen2010}Christen, J. A. and Fox, C. A general purpose sampling algorithm for continuous distributions (the t-walk). {\em Bayesian Anal.} \textbf{5} (2) 263 - 281 https://doi.org/10.1214/10-BA603
\bibitem[Debras \& Chabrier(2019)]{de19} Debras, F. \& Chabrier, G.\ 2019, \apj, 872, 100. 
\bibitem[Duer et al.(2020)]{Duer2020}{Duer}, K., {Galanti}, E., and {Kaspi}, Y.\ 2020, J. Geophys. Res. (Planets), 125, 8.
\bibitem[Durante et al.(2020)]{du20}Durante, D., Parisi, M., Serra, D., et al.\ 2020, Geophysical Research Letters, 47, e86572.
\bibitem[Fletcher et al.(2020)]{Fletcher+2020}{Fletcher}, L.~N. et al. \ 2020, Journal of Geophysical Research, 125, 8. 
\bibitem[Galanti et al.(2016)]{Galanti2016}{Galanti}, E. and {Kaspi}, Y. \ 2016, ApJ, 820, 91.
\bibitem[Galanti et al.(2017)]{Galanti2017b}Galanti, E. and Kaspi, Y. \ 2017, Icarus, 286, 46--55.
\bibitem[Galanti et al.(2019)]{Galanti2019a}{Galanti}, E., {Kaspi}, Y., {Miguel}, Y., {Guillot}, T., {Durante}, D., {Racioppa}, P., and {Iess}, L. \ 2019, Geophys. Res. Lett., 46, 2.
\bibitem[Goodman \& Weare(2010)]{Goodman2010} Goodman, J. and Weare, J.\ 2010, Communications in Applied Mathematics and Computational Science, 5, 65–80
\bibitem[Guilera et al.(2020)]{octi20} Guilera, O.~M., S{\'a}ndor, Z., Ronco, M.~P., et al.\ 2020, \aap, 642, A140. 
\bibitem[Guillot \& Morel(1995)]{cepam95}Guillot, T., \& Morel, P. 1995, \aap Supplement Series, 109, 109-123.
\bibitem[Guillot et al.(2018)]{gu18}Guillot, T. et al. \ 2018, Nature, 555, 227.
\bibitem[Guillot et al.(2020)]{Guillot+2020b}{Guillot}, T., et al., \ 2020, Journal of Geophysical Research, 125, 8.
\bibitem[Gupta et al.(2022)]{gu22}{Gupta}, T., et al., \ 2022, in preparation.
\bibitem[Hubbard (2013)]{hu13}Hubbard, W. B. \ 2013, ApJ, 768, 43.
\bibitem[Hubbard \& Militzer(2016)]{hu16} Hubbard, W.~B. \& Militzer, B.\ 2016, \apj, 820, 80. 
\bibitem[Iess et al.(2018)]{ie18}Iess, L. et al. \ 2018, Nature, 555, 220–222.
\bibitem[Kaspi(2013)]{Kaspi2013} Kaspi, Y.\ 2013, \grl, 40, 676. 
\bibitem[Kaspi et al.(2016)]{Kaspi2016}{Kaspi}, Y., {Davighi}, J.~E., {Galanti}, E., and {Hubbard}, W.~B. \ 2016, Icarus, 276, :170--181. 
\bibitem[Kaspi et al.(2017)]{Kaspi2017}Kaspi, Y. et al. \ 2017, Geophys. Res. Lett., 44, 12.
\bibitem[Kaspi et al.(2018)]{kaspi2018}Kaspi, Y., et al. \ 2018, Nature, 555, 223--226.
\bibitem[Kaspi et al.(2020)]{kaspi2020} Kaspi, Y., Galanti, E., Showman, A.~P., et al.\ 2020, \ssr, 216, 84. 
\bibitem[Kulowski et al.(2020)]{kulowski20} Kulowski, L., Cao, H., \& Bloxham, J. \ 2020, Journal of Geophysical Research: Planets, 125.
\bibitem[Lambrechts et al.(2014)]{la14} Lambrechts, M., Johansen, A., \& Morbidelli, A.\ 2014, \aap, 572, A35. 
\bibitem[Leconte et al.(2017)]{Leconte+2017}{Leconte}, J., {Selsis}, F., {Hersant}, F., {Guillot}, T. \ 2017, 598, A98.
\bibitem[Li et al.(2020)]{li20}Li, C. et al. \ 2020, Nature Astronomy, 4, 609.
\bibitem[Lindal(1981)]{Lindal1981}{Lindal}, G.~F. et al. \ 1981, Journal of Geophysical Research, 86, 8721–8727.
\bibitem[Lyon \& Johnson(1992)]{SESAME}Lyon, S., \& Johnson, J. , 1992, LANL Report LA-UR-92-3407, Los Alamos
\bibitem[Mazevet et al.(2021)]{mazevet20}{Mazevet}, S. and {Licari}, A. and {Soubiran}, F.\ 2021 \textit{arXiv:2012.09454}.
\bibitem[Miguel et al.(2016)]{mi16}Miguel, Y., Guillot, T., \& Fayon, L.\ 2016, \aap, 596, A114.
\bibitem[Militzer \& Hubbard(2013)]{MH13}Militzer B. \& Hubbard W. B. \ 2013, ApJ, 774, 148.
\bibitem[Morales et al.(2013)]{mo13}Morales, M. A., Hamel, S., Caspersen, K. \& Schwegler, E. \ 2013, Physical Review B, 87, 174105.
\bibitem[Nettelmann (2017)]{n17}Nettelmann, N. \ 2017, \aap, 606, A139.
\bibitem[Nettelmann et al.(2021)]{ne21} Nettelmann, N., Movshovitz, N., Ni, D., et al.\ 2021, The Planetary Science Journal, 2, 241. 
\bibitem[Ni(2019)]{Ni19} Ni, D.\ 2019, \aap, 632, A76. doi:10.1051/0004-6361/201935938
\bibitem[Ormel et al.(2021)]{ormel21}Ormel, C.~W., Vazan, A., \& Brouwers, M.~G.\ 2021, \aap, 647, A175.
\bibitem[Pollack et al.(1996)]{po96} Pollack, J.~B., Hubickyj, O., Bodenheimer, P., et al.\ 1996, \icarus, 124, 62.
\bibitem[Saumon et al.(1995)]{SCvH95}Saumon, D., Chabrier, G. \& van Horn, H. M. \ 1995, ApJ Supplement Series, 99, 713.
\bibitem[Seiff, A. et al.(1998)]{se98} Seiff, A. et al. \ 1998, Journal of Geophysical Research, 103 , 22857–22890.
\bibitem[Serenelli et al.(2010)]{sa10}Serenelli, A. M. \& Basu, S. \ 2010, ApJ, 719, 865.
\bibitem[Sch{\"o}ttler et al.(2018)]{scr18}Sch{\"o}ttler, M., Redmer, R.\ 2018, Physical Review Letters, 120.
\bibitem[Tollefson et al.(2017)]{Tollefson2017}Tollefson, J. et al. \ 2017, Icarus, 296 , 163--178.
\bibitem[Valletta \& Helled (2019)]{vh19} Valletta, C. \& Helled, R.\ 2019, ApJ, 871, 127.
\bibitem[Vazan et al.(2018)]{va18}Vazan, A., Helled, R., \& Guillot, 2018, \aap, 610, L14.
\bibitem[Venturini \& Helled (2020)]{ve20}Venturini, J., Helled, R.\ 2020, \aap, 634, A31.
\bibitem[Wahl et al.(2017)]{wa17} Wahl, S. M. et al. \ 2017, Geophys. Res. Lett., 44, 4649.
\bibitem[von Zahn et al.(1998)]{vo98}von Zahn, U., Hunten, D. M., \& Lehmacher, G. \ 1998, Journal of Geophysical Research, 103, 22815.
\bibitem[Zharkov \& Trubitsyn (1978)]{zt78}Zharkov, V. N. \& Trubitsyn, V. P.\ Physics of planetary interiors .\textit{Astronomy and Astrophysics Series, Tucson: Pachart} (1978). 
\end{thebibliography}
\end{document}